\documentclass[twocolumn,preprintnumbers,aps]{revtex4}
\usepackage{graphicx} 
\begin{document}

\preprint{ For Physical Review Letters }

\title{ Absence of correlation between built-in electric dipole moment and quantum Stark
effect in InAs/GaAs self-assembled quantum dots }

\author{ Weidong Sheng and Jean-Pierre Leburton }

\affiliation{Beckman Institute for Advanced Science and Technology and Department of
Electrical and Computer Engineering, University of Illinois at Urbana-Champaign, Urbana,
Illinois 61801}

\begin{abstract} 
We report significant deviations from the usual quadratic dependence of the ground state
interband transition energy on applied electric fields in InAs/GaAs self-assembled quantum
dots. In particular, we show that conventional second-order perturbation theory fails to
correctly describe the Stark shift for electric field below $F = 10$~kV/cm in high dots.
Eight-band ${\bf k}\cdot{\bf p}$ calculations demonstrate this effect is predominantly due to
the three-dimensional strain field distribution which for various dot shapes and
stoichiometric compositions drastically affects the hole ground state. Our conclusions are
supported by two independent experiments.
\\PACS numbers: 78.67.Hc, 73.21.La, 31.15.Md
\end{abstract}

\maketitle

Self-assembled InAs/GaAs quantum dots (SADs) are three-dimensional (3D) semiconductor
nanostructures in which electrons and holes are completely confined along the three dimensions
of space by the band gap difference between InAs and GaAs materials \cite{bim}. In these
nanoscale systems, the determination of the electronic spectra of both particles represents a
major challenge because of the low symmetry of the 3D confined nanostructures that take
(truncated) pyramidal or lens shapes, and are affected by the strong influence of strain due
to the lattice mismatch between the InAs and GaAs crystals \cite{keg}. Experimentally, the
difficulty of ascertaining the exact dot shape, and the non-uniformity in size distribution
resulting from the growth process are also a handicap \cite{ebi}. In this context, the
knowledge of the respective positions of electrons and holes can provide information on the
confining potential experienced by both particles, which is of primary importance for
fundamental \cite{gam} as well as practical reasons \cite{dep}.

It has been recently argued that because of the non-homogenous stoichiometric composition of
SADs, resulting from atomic InAs-GaAs inter-diffusion, the center of mass of the ground state
electrons and holes are displaced from one another with the hole above the electron, thereby
inducing a built-in electric dipole oriented from the top to the base of the dot. This
inverted electron-hole alignment \cite{fry} has been derived experimentally by assuming the
usual linear relation between the electric dipole moment and the Stark shift in Stark effect
spectroscopy on SADs in p-i-n structures, {\it i.e.}, \begin{equation} E(F) = E_0 + p F +
\beta F^2 \end{equation} where $E$ is the electron-hole ground state transition energy, $p$ is
the built-in dipole moment, and $\beta$ measures the polarization of the electron and hole
states.

By estimating the electric field at which the maximum transition energy occurs, one can
determine whether the structure has a positive (if that happens at positive fields) or
negative (if otherwise) built-in dipole moment. This quadratic dependence is well-known for
quantum-well systems and has been confirmed in nanocrystallite quantum dots \cite{emp}. Recent
theoretical works have made use of this linear relation to support the universality of the
inverted alignment in SADs \cite{bak}.

\begin{figure}
\includegraphics[width=3.5in]{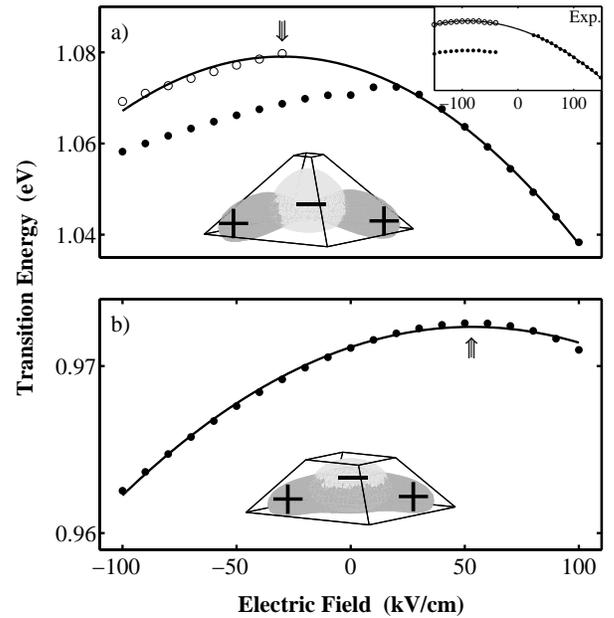}
\caption{ Ground state interband transition energies as a function of electric fields for two
truncated pyramidal self-assembled quantum dots. Probability density isosurfaces of the ground
states of electron (dark grey) and hole (light grey) are shown on a schematic view of the
structure. The calculated points are shown in dotted lines. a): The solid line is a fit to
Eq.~1 using only the data at positive electric fields. The data at negative fields are
shifted to the open dots by $11$~meV to match the fitted curve. The top-right inset shows the
corresponding experimental data taken from Ref. \cite{fry}. b): The solid line is a parabolic
fit to all the data. }
\end{figure}

In this letter, we demonstrate that the linear relation between electric dipole moment and
Stark shift as expressed in Eq.~1 is usually violated in SADs because of the unique strain
field distribution in the dot, which strongly influences the rapidly varying confining
potential for the holes. We show that this peculiar effect invalidates the conventional
perturbation approach to establish the electric field dependence of the optical transition in
SADs.

The system of SAD structures, containing both electrons and holes, can be well described by
an eight-band strain-dependent ${\bf k}\cdot{\bf p}$ Hamiltonian \cite{bah}, which
reads,
\begin{equation}
({\bf H}_{k\cdot p}^0 + {\bf H}_{k\cdot p}^s + |e|Fz) \psi = E \psi.
\label{eq:H} 
\end{equation}
Here ${\bf H}_{k\cdot p}^0$ and ${\bf H}_{k\cdot p}^s$ is the kinetic and strain-dependent
part of the eight-band Hamiltonian \cite{pry}, respectively, and $\psi=(\psi_1, \psi_2,
\ldots, \psi_8)$ is the envelop eigenvector. By using continuum elasticity theory, the strain
tensor is calculated on a large grid of $150\times 150\times 150$ sites. The Hamiltonian is
then solved by Lanczos algorithm. The same technique has been successfully applied to study
few-particle effect \cite{lan} and optical transitions \cite{swd2} in quantum dots and very
good agreement with experiments has been achieved.

Figure 1 shows the calculated dependence of interband transition energies on electric fields
for two SAD structures. Fig.~1(a) illustrates the data for a $\mbox{In}_{0.8} \mbox{Ga}_{0.2}
\mbox{As}$ dot with a constant composition throughout the structure. The pyramidal dot is
$18$~nm wide and $7.8$~nm high, which is of similar size as that in the experiment of Fry {\it
et al.} \cite{fry}. The electric field is applied vertically to the structure, pointing from
the base to the top. For comparison, the corresponding experimental data is shown in the
top-right inset.

An clear deviation from the quadratic dependence of the transition energy on the electric
field is obtained from our numerical simulations that show two distinct branches, one for
roughly each field direction, merging at around zero field. It is interesting to notice that,
if the data at negative electric fields are shifted upwards by $11$~meV, and if one ignores a
few data points around zero-field, one can easily fit the remaining data with a single
parabola characterized by a maximum at around $F=-40$~kV/cm. In this case, according to
Eq.~1, the displaced curve yields a negative built-in dipole ($-14$~\AA) in spite of the
fact that the center of mass of the electron state is above that of the hole state by $7$~\AA~
as shown on the lower inset, thereby achieving normal alignment.

This fitting procedure was adopted by Fry {\it et al.} \cite{fry} to show the overall
quadratic field dependence of the experimental transition energies (see upper right inset).
The points marked with solid dots are the data measured from experiments. We notice however
the structure may be of different stoichiometric composition than our SAD in Fig.~1(a). The
authors shifted their data to the position marked with open dots to match the curve fitting
from their data at positive electric fields. With this procedure, they claimed negative
built-in dipole in the quantum dot structure. This experimental procedure was justified by the
fact that for p-i-n structures, the Stark effect can only be measured under reverse biases.
Therefore, given the asymmetric SAD shape and the fact that the Stark effect was measured for
both field directions, two sets of samples i.e p-i-n and n-i-p structures, were used in the
experiment.

\begin{figure}
\includegraphics[width=3.5in]{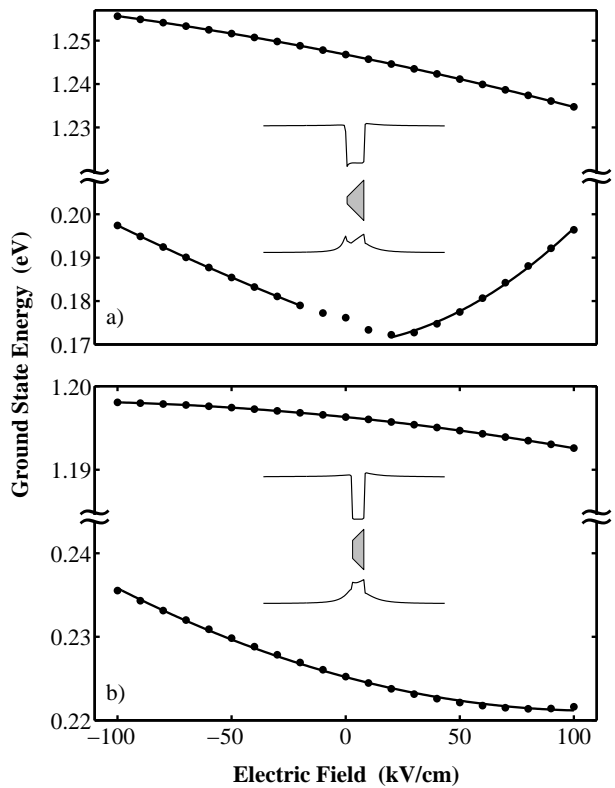}
\caption{ Ground state energies of electrons and holes as a function of electric fields for
the two quantum dots shown in Fig.~1. The band diagrams are shown in the insets. }
\end{figure}

While the conventional perturbation theory fails for the pyramidal quantum dot, it makes a
successful prediction of Stark shifts for another quantum dot which is a largely truncated
pyramidal InAs quantum dot shown in Fig.~1(b), which has the same base size, but is only
$5.4$~nm high. It has smaller transition energies than the pyramidal structure because of its
uniform InAs composition. Unlike the dot shown in Fig.~1(a), it has nearly perfect quadratic
dependence at electric fields as strong as $F=\pm 100$kV/cm. The parameter extracted from the
fitting curve gives a positive built-in dipole of $4.57$~$\mbox{\AA}$ which agrees well with
the actual value $4.8$~\AA.

In Fig.~2, we show the electron and hole ground state energies for the two structures as a
function of electric fields. Electron energies for both structures are seen to have a linear
profile with a slight bowing that can be well described by the usual quadratic dependence on
the electric field, {\it i.e.}, \begin{equation} E_e(F) = E_e(0) + p_e F + \beta_e F^2,
\end{equation} in the second-order perturbation theory. The hole state in the truncated
structure exhibits a similar feature and its energies can be well fitted with a parabola, {\it
i.e.}, \begin{equation} E_h(F) = E_h(0) + p_h F + \beta_h F^2. \end{equation} Here
$p_{e(h)}/|e| = \langle \psi_{e(h)}|z|\psi_{e(h)} \rangle$ is the center of mass of the
electron (hole) state $\psi_e$ ($\psi_h$) and $\beta_{e(h)}$ measures the polarization of the
electron(hole) states. The correlation between the built-in dipole moment $p=p_e-p_h$ and
quantum Stark effect is then obtained by taking the difference between the electron and hole
energies to reflect Eq.~1. In the pyramidal dot, however, a quadratic fitting for the whole
field range is not possible, although the left and right branches of the spectrum could be
fitted piecewise by two parabolas, respectively.

The different behavior of electron and hole energies is attributed to their respective band
edge profiles shown in the insets of Fig.~2. In both SADs, the conduction band edge is almost
flat inside the dot with hard walls on the sides, which provides electrons a constant, strong
confinement. Consequently, under the influence of external electric fields, electrons
experience a smoothly varying confining potential inside the dot, which can be treated as a
small perturbation with electron energies following the quadratic dependence given by Eq.~3.
This type of behavior is also observed for holes in the truncated SAD structure (Eq.~4).

In the pyramidal dot, the valence band edge exhibits a more complicated, double-triangular
profile with the lower part extending from the SAD base to about two-third of the structure
hight, which effectively confines the ground hole state close to the bottom of the dot (see
the inset in Fig.~1). Therefore, the holes are localized in a rapidly varying potential while
the valence band edge changes abruptly close to the top of the structure. Hence, as vertical
electric fields force the holes to move along the structure, they experience significantly
different local confinement at different fields, which invalidates the perturbation approach
as described in Eq.~1. This behavior is seen to occur at electric fields as low as $F=\pm
10$~kV/cm.

\begin{figure}
\includegraphics[width=3.5in]{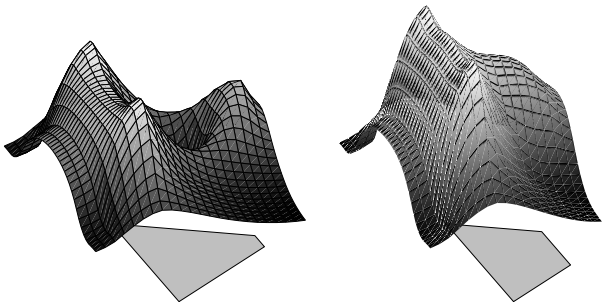}
\caption{ Three-dimensional plots of strain distribution in a [100] plane across the center of
quantum dots. Below the plots are the schematic view of cross sections of the structures. }
\end{figure}

The rapidly varying confining potential which is responsible for the breakdown of the
perturbation theory in the pyramidal structure is induced by the unique 3D strain field
distribution within the SAD. Because the effect of other strain components on the valence
bands are inferior in importance, we concentrate on the biaxial component of the strain which
affects mostly the valence bands \cite{bah2}. Fig.~3 shows a 3D plot of the magnitude of
biaxial strain along a [100] plane across the center of the quantum dots. It is pointed out
that the biaxial component of strain takes negative value inside the dot that is under biaxial
compression and positive value in the barrier that is under biaxial tension.

Despite the fact that the structural difference between the two dots results only from the
truncation, both structures exhibit substantially different strain distribution. The biaxial
strain distribution in the pyramidal dot shows rapid variation inside the structure along the
growth direction while the truncated dot exhibits a smoother profile. The strain field is so
coherent that the tip of the pyramidal dot has a significant effect on the whole strain
distribution.

\begin{figure}
\includegraphics[width=3.5in]{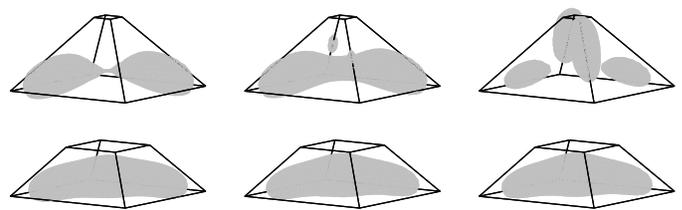}
\caption{ Probability density isosurfaces of ground hole states at different electric fields,
from left to right, $-20$~kV/cm, $10$~kV/cm, and $20$~kV/cm. Top panel: pyramidal dot
(Fig.~1a); Bottom panel: truncated dot (Fig.~1b). }
\end{figure}

In Fig.~4, we show the ground hole states at both negative and positive electric fields for
the two structures. For the pyramidal structure of Fig.~1(a), the wave function at negative
field $F=-20$~kV/cm looks similar to that at zeros field except that it is localized closer to
the dot bottom. At a positive field $F=10$~kV/cm, the hole wave function develops a pair of
small `wings' around the facet edges \cite{pry2}. Because of a strong local confinement
imposed by the sharp triangular potential (see the band diagram inset in Fig.~2), the hole
state could not extend into the upper half part of the structure as electron states do. At a
stronger positive field $F=20$~kV/cm, these wings even dominate over the other portions of the
wave function.

The obvious asymmetric behavior of the ground hole states implies that the corresponding wave
functions cannot be expanded to the first order in the electric field as \begin{equation}
\psi_h(F) = \psi_h(0) + \phi_h F, \end{equation} which would result into symmetric wave
functions with respect to positive and negative electric fields for the whole range of
electric fields. Instead, if we chose different values of the perturbation, $\phi_h$, for
negative and positive fields, respectively, it is possible to fit the hole energies with two
parabolas (see Fig.~2).

In the truncated structure, the hole states are seen to have a different probability density
distribution. The wave functions change very smoothly with the electric fields, because of a
relatively flat valence band edge (see Fig.~2 inset). In addition, the lateral confinement
close to the SAD top is also weaker compared with that in the pyramidal SAD as shown Fig.~1.
Therefore, hole states behave coherently throughout the whole range of electric fields,
exhibiting nearly perfect quadratic dependence of their eigen energies on electric fields.

\begin{table}
\caption{ Calculated built-in dipole moments and their fitted values from the assumed
quadratic dependence on the field for various quantum dot structures. All the lengths are
expressed in nm. If not specified, all structures are homogeneous InAs/GaAs dots. }
\begin{ruledtabular}
\begin{tabular}{ccc}
Size & Actual value & Fitted value \\
$18\times 7.8$\footnotemark[1] & -0.31 & N/A \\
$18\times 7.8$\footnotemark[2] & 0.70 & -1.40 \\
$18\times 7.8$                 & 0.33 & -1.23 \\
$18\times 9.0$                 & 0.36 & -1.33 \\
$18\times 6.6$                 & 0.54 & 0.39  \\
$18\times 5.4$                 & 0.46 & 0.48  \\
\end{tabular}
\end{ruledtabular}
\footnotetext[1]{Inhomogeneous InAs dot with two interdiffused monolayers, $\mbox{In}_{0.6}
\mbox{Ga}_{0.4} \mbox{As}$ and $\mbox{In}_{0.8} \mbox{Ga}_{0.2} \mbox{As}$, in the bottom.}
\footnotetext[2]{The $\mbox{In}_{0.8}\mbox{Ga}_{0.2}\mbox{As}$ dot shown in Fig.~1(a).}
\end{table}

Table~I lists the actual and fitted values of dipole moment for several SADs. All structures
have the same base dimensions, but different heights or composition profiles. It should be
stressed that the fitted values are obtained from the model data only for positive electric
fields, {\it i.e.}, the right branch of the spectrum. Except for the last structure, all
values exhibit large discrepancy between the fitted curve and the model (calculated) data at
negative fields. The first structure has a negative built-in dipole induced by the
inhomogeneous diffusion \cite{swd}, responsible for two bumps in the transition energy for
positive and negative fields, which makes the fitting over a broad field range, meaningless.
This structure is similar to the SADs investigated recently by Chen {\it et al.} \cite{mad}
where a negative dipole moment was extracted (the orientation of their field was opposite to
ours) from a quadratic fitting over a smaller range of field ($-7.5$~kV/cm~$\leq F
\leq$~$10$~kV/cm), in good agreement with our data. The next three structures have homogeneous
composition, and all exhibit positive dipoles. However, the fitted values for these structures
are all negative. Therefore, it is a general feature of high pyramidal quantum dots with less
than $25\%$ truncation to exhibit erroneously a 'negative' built-in dipole when fitted to the
Stark shifts. The fourth structure has about $27\%$ truncation with a dipole larger than those
in the two previous dots. The corresponding fitted value is positive but is $30\%$ smalled
than the actual value. In the last structure, which is strongly truncated, the difference
between the actual value and the fitted one becomes negligible.

In conclusion, we have shown that the unique 3D strain field distribution is responsible for
the breakdown of the linear dependence of the built-in electric dipole moment in the Stark
shift in SADs. This effect is mostly noticeable in high SADs where holes are mostly affected
by rapidly varying components of the strain induced confining potential, which invalidates the
second order perturbation theory. While perturbation theory is expected to fail at strong
electric fields, we point out that significant deviations from the usual Stark effect already
occur at fields lower than $F=10$~kV/cm. For largely truncated or flat SADs, the Stark shift
retains its usual quadratic dependence on external fields.

\acknowledgments
This work is supported by Army Research Office, grant \#DAAD 10-99-10129 and National
Computational Science Alliance, grant \#ECS000002N.

\end{document}